\pdfoutput=1
\documentclass[12,preprint]{aastex}
\usepackage{graphicx}
\def\etal{{\rm et~al.}}
\title{MEASURING THE HUBBLE CONSTANT WITH THE HUBBLE SPACE TELESCOPE\footnote{Based on observations made with the NASA/ESA Hubble
Space Telescope obtained at the Space Telescope Science Institute which is operated by AURA for NASA.}} 

\author{W.L. Freedman, R.C. Kennicutt, \& J.R. Mould}           

\begin{document}

\maketitle

\abstract
Ten years ago our team completed the Hubble Space Telescope
Key Project on the extragalactic distance scale. Cepheids
were detected in some 25 galaxies and used to calibrate four
secondary distance indicators that reach out into the expansion
field beyond the noise of galaxy peculiar velocities. The result
was H$_0$ = 72 $\pm$ 8 km s$^{-1}$ Mpc$^{-1}$ and put an end to
galaxy distances uncertain by a factor of two. This work has
been awarded the Gruber Prize in Cosmology for 2009.
\endabstract

\section{Introduction}
\noindent
Our story begins in the mid-1980s with community-wide discussions on Key Projects
for the NASA/ESA Hubble Space Telescope. The extragalactic distance scale was the
perfect example of a project that would be awarded a generous allocation of time
in order that vital scientific goals would be accomplished even if the lifetime of
the telescope proved to be short. As Marc Aaronson (Figure 1), the original principal
investigator of the project, said in his Pierce Prize Lecture in 1985, ``The distance scale 
path has been a long and torturous one, but with the imminent launch of HST there is 
good reason to believe that the end is finally in sight."

\begin{figure}[h]

\includegraphics[width=7.5cm]{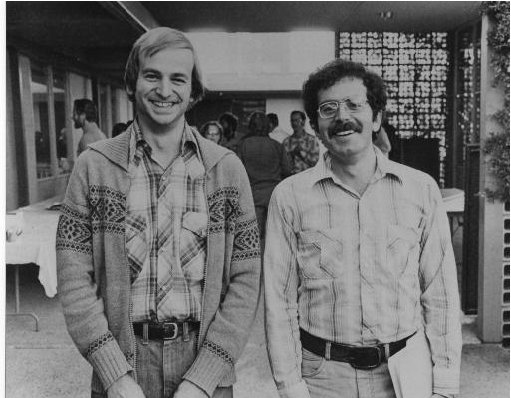}
\caption{Jeremy Mould (left) and Marc Aaronson (right) at the van Biesbroeck Prize
award ceremony in 1981.}
\end{figure}

Marc Aaronson died tragically in an accident in 1987, having written a successful
proposal for the Key Project, a project designed to shrink the scatter shown in
Figure 2 to 10\% and put an end to 60 years of debate, commencing with Hubble's
estimates in 1929.

\begin{figure}[ht]
\includegraphics[width=15cm]{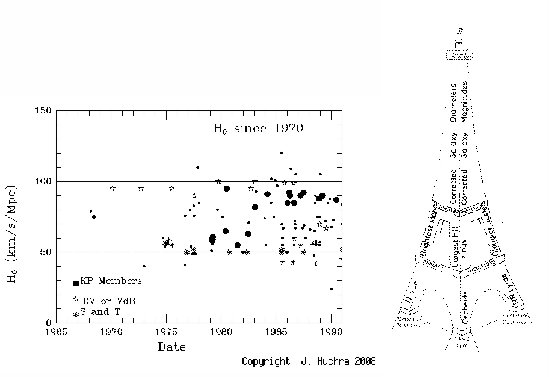}
\caption{The scatter in estimates of the Hubble Constant showed no sign of convergence
in the 1970s and 1980s. On the right is a schematic distance scale by G. de
Vaucouleurs, one of the protagonists in the distance scale controversy.}
\end{figure}

The principal reason for the uncertainty in H$_0$ is evident in Figure 3. Large 
scale structure is seen out to distances of 100 Mpc. Ground-based Cepheid distances,
however, extended to only 4 Mpc with a ``twilight zone" beyond (Figure 4). 

\begin{figure}[ht]
\includegraphics[width=10cm]{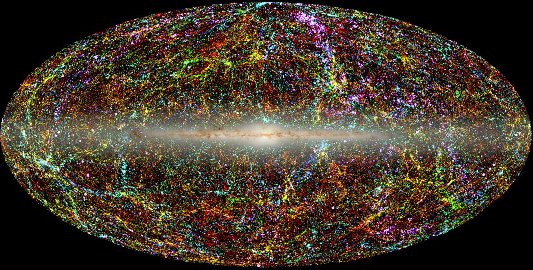}
\caption{Colour coded redshifts show the large scale structure in the nearby Universe
(galactic coordinates).}
\end{figure}

The Key Project's solution to the twilight zone problem was to map Cepheids out to
20 Mpc and calibrate secondary distance indicators within this volume. The secondary
indicators would extend the distance scale out to 100 Mpc.

\begin{figure}[ht]
\includegraphics[width=15cm]{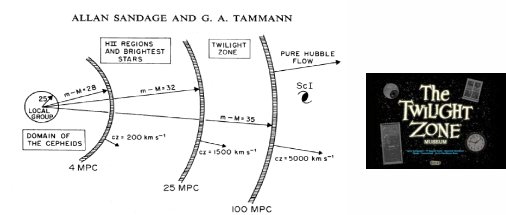}
\caption{A schematic distance scale by other protagonists of the controversy,
A. Sandage \& G. Tammann.}
\end{figure}

An important tenet of the Key Project was to exploit the redundancy of distance
indicators, as shown in Figure 5, especially the four secondary distance indicators,
the Tully-Fisher relation, surface brightness fluctuations, supernovae, and the
fundamental plane.

\begin{figure}[ht]
\includegraphics[width=10cm]{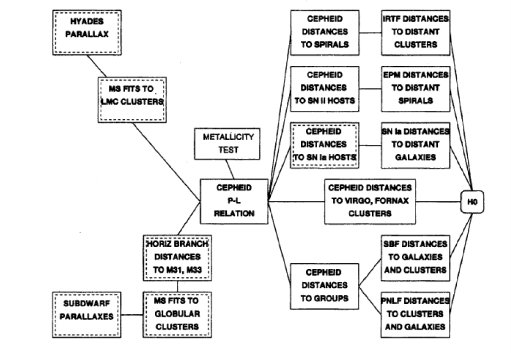}
\caption{The schematic distance scale in the Key Project Committee report.}
\end{figure}

\section{Cepheids}
\noindent
The Cepheid Period-Luminosity law was discovered by Leavitt early in the twentieth
century.

\begin{figure}[ht]
\includegraphics[width=10cm]{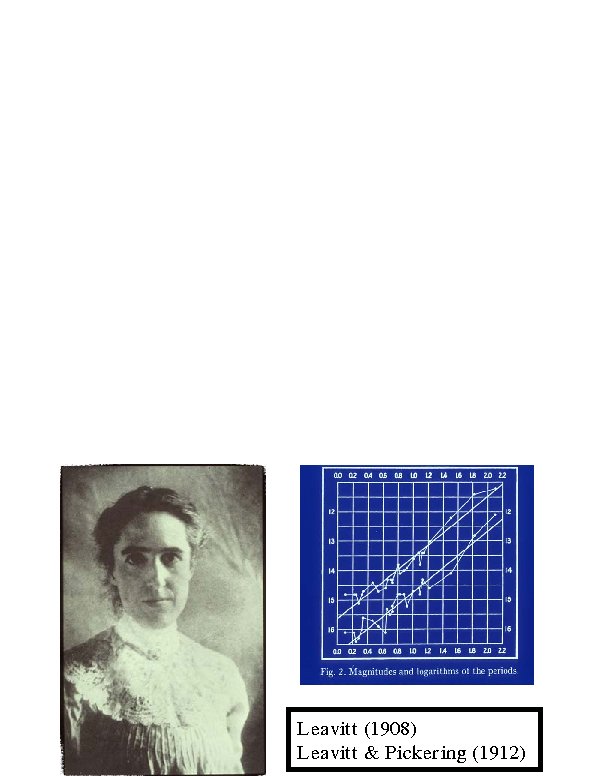}
\caption{Henrietta Leavitt and the period-luminosity relation in the Magellanic Clouds.}
\end{figure}

By the end of the century much was understood about the systematics of Cepheids.
For example (Madore \& Freedman 1991), Cepheid amplitudes are maximum in the blue,
and interstellar absorption is minimum at long wavelengths (Figure 7). Therefore, the best
strategy for discovering Cepheids is to observe at visible wavelengths; to minimize
the effect of dust luminosities are best measured in the infrared. 

\begin{figure}[ht]
\includegraphics[width=5cm]{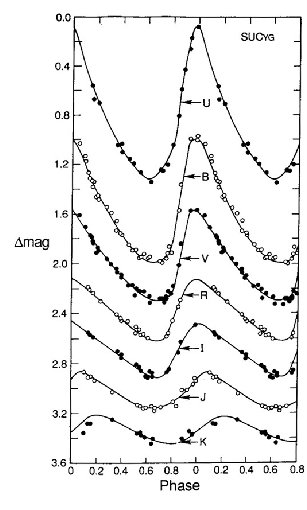}
\caption{Amplitude as a function of wavelength for Cepheid variables.}
\end{figure}

HST brought a number of strengths to the Cepheid distance scale: linear detectors,
multiwavelength observations, and a planned cadence of observations. Figure 8
shows for M33 (Freedman et al 1991) how an absolute distance modulus is obtained
with knowledge of the reddening law.

\begin{figure}[ht]
\includegraphics[width=5cm]{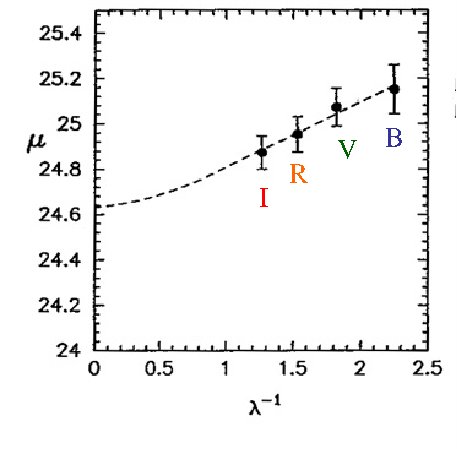}
\caption{Finding the distance of M33 (Freedman et al 1991).}
\end{figure}

Figure 9 shows how a power law observing cadence is superior for luminosity 
measurement to equally spaced observations (Madore \& Freedman 2002, 2005).

\begin{figure}[ht]
\includegraphics[width=5cm]{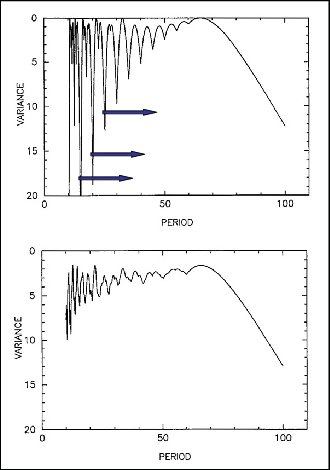}
\caption{Observing cadence. Upper panel: equally spaced observations.
Lower panel: power law cadence. }
\end{figure}

The results are shown in Figure 10. The light curves for these periods
are unmistakably Cepheids. Our observations populated the range 10--100
days in galaxies with distances of order 10 Mpc.

\begin{figure}[ht]
\includegraphics[width=12.5cm]{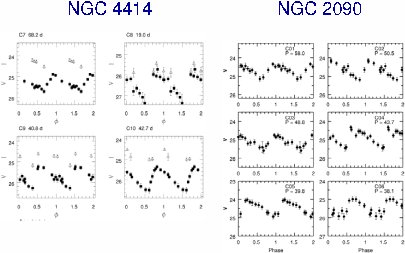}
\caption{Light curves for Cepheids in NGC 2090 (Phelps et al 1998) and
NGC 4414 (Turner et al 1998). }
\end{figure}

Figure 11 shows a typical field placement for the Key Project in the
large face-on spiral galaxy in the Virgo cluster, M100. Twelve V observations
were obtained during the sequence and four I observations. Cosmic ray splits
were used and a fixed roll angle was adopted. Photometry was carried out on
the frames using DoPhot and a custom version of DAOPHOT called ALLFRAME
(Stetson 1994).

\begin{figure}[ht]
\includegraphics[width=5cm]{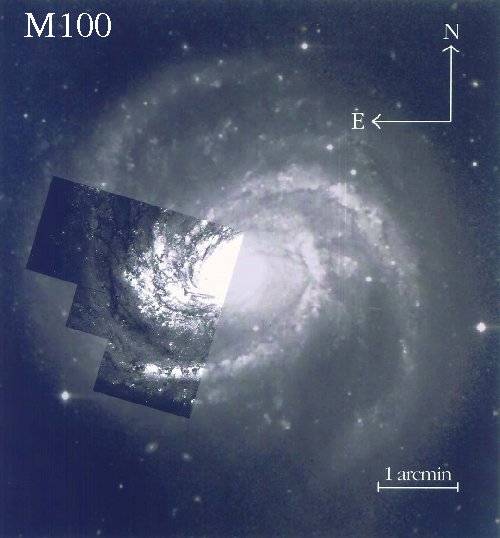}
\caption{The WFPC2 footprint is shown on an image of M100. }
\end{figure}

A composite I-band period-luminosity relation for 800 Cepheids in 24 galaxies is shown in
Figure 12 (corrected for distance) (Ferrarese et al 2000).

\begin{figure}[ht]
\includegraphics[width=10cm]{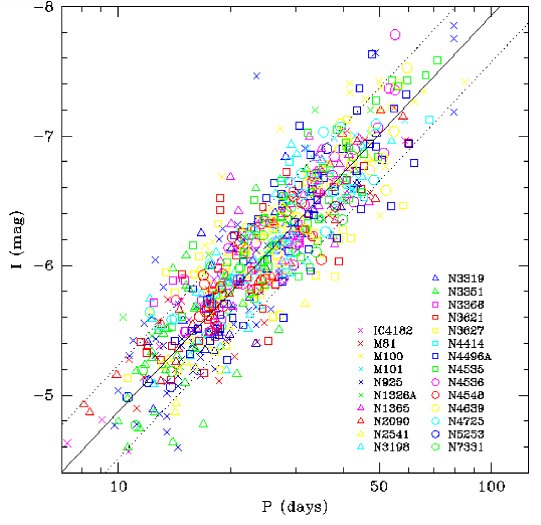}
\caption{Composite I-band PL relation for the Key Project. }
\end{figure}

Ferrarese \etal~ (2000) also carried out a comprehensive comparison of Key Project
distances with other prominent distance indicators, such as the tip of the red giant
branch (e.g. Sakai \etal~ 1997) and surface brightness fluctuations (Tonry \etal~ 2001)
(Figure 13) and the planetary nebula luminosity function (Ciardullo \& Jacoby 1992) and the globular
cluster luminosity function (Secker \& Harris 1993) (Figure 14).

\begin{figure}[ht]
\includegraphics[width=15cm]{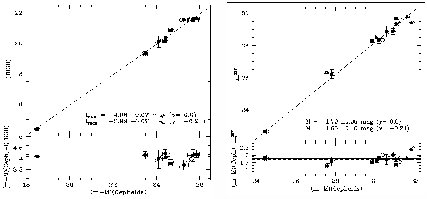}
\caption{Comparison of Cepheid distances with the tip of the red giant (left)
and surface brightness fluctuations (right). }
\end{figure}

\begin{figure}[ht]
\includegraphics[width=15cm]{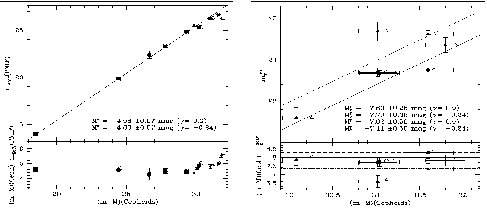}
\caption{Comparison of Cepheid distances with the planetary nebula luminosity function (left)
and the globular cluster luminosity function (right). }
\end{figure}

\section{Metallicity Calibration}
\noindent
The Cepheid period luminosity relation is a straightforward primary distance indicator
if the precepts described above are followed. But there is a complication. Cepheids
vary in their chemical composition, and the period luminosity relation is affected.
We can write

$$(m-M)_{true} = (m-M)_{PL} - \gamma \log(Z/Z_{LMC})$$

where $Z$ is the metallicity of the field and $Z_{LMC}$ is the metallicity
of the LMC.

However, theory is not predictive, even about the sign of $\gamma$. According to
Chiosi, Wood \& Capitanio (1993) $\gamma_{VI}$ = --0.11 mag/dex, but elsewhere we
find with different opacities $\gamma_{VI}$ = +0.06 mag/dex. The Key Project took
an empirical approach, described by Kennicutt (1998). Both an inner and an outer
field were observed in M101 (Figure 15), a galaxy with a large abundance gradient
(Figure 16).
The Cepheid metallicities were assumed to follow the oxygen abundances of nearby
HII regions. The inner field modulus was found to be 29.20 $\pm$ 0.09 mag, and the
outer modulus was 29.36 $\pm$ 0.08 mag. There was a factor of five difference in Z.
This yielded $\gamma$ = --0.24 $\pm$ 0.16. Oxygen abundances were measured for each
of the Key Project galaxies by Zaritsky, Kennicutt, and Huchra (1994).

\begin{figure}[ht]
\includegraphics[width=10cm]{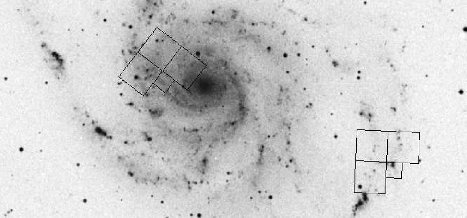}
\caption{The two WFPC2 fields observed in M101.}
\end{figure}

\begin{figure}[ht]
\includegraphics[width=7.5cm]{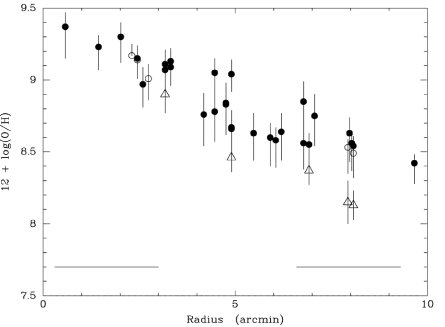}
\caption{The gradient in oxygen abundance in M101.}
\end{figure}

Confirmation of this result comes from a comparison of tip of the red giant branch
distances (TRGB) and Cepheid distances by Ferrarese \etal~ (2000) (Figure 17).

\begin{figure}[ht]
\includegraphics[width=7.5cm]{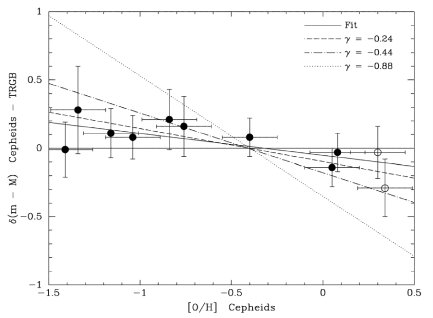}
\caption{Comparison of TRGB and Cepheid distances for different values of $\gamma$.}
\end{figure}

More recent work has strengthened these results. Mould \& Sakai (2008, 2009ab)
have shown that substitution of TRGB distances for Cepheid distances in 
secondary distance indicator calibrations returns a Hubble Constant in agreement
with the Key Project. And Scowcroft \etal~ (2009) obtained $\gamma_{VI}$ = --0.26
in a study of M33. 

\section{Measurement of the Hubble Constant}
\noindent
Four secondary distance indicators were calibrated by the Key Project. The first
was the Tully-Fisher relation (Figure 18). Sakai \etal~ (2000) obtained H$_0$ =
71 $\pm$ 3 $\pm$ 7 km s$^{-1}$ Mpc$^{-1}$, where the first uncertainty is the
random error and the second uncertainty is the systematic error.

\begin{figure}[ht]
\includegraphics[width=15cm]{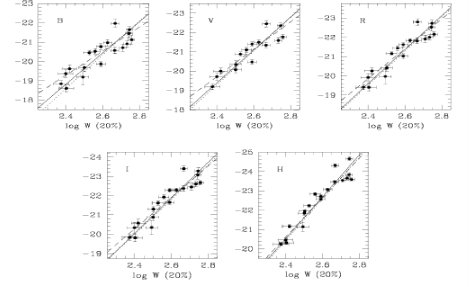}
\caption{Distances to 21 Cepheid galaxies calibrate the Tully-Fisher relation in 
the BVRIH bandpasses.}
\end{figure}

\begin{figure}[ht]
\includegraphics[width=10cm]{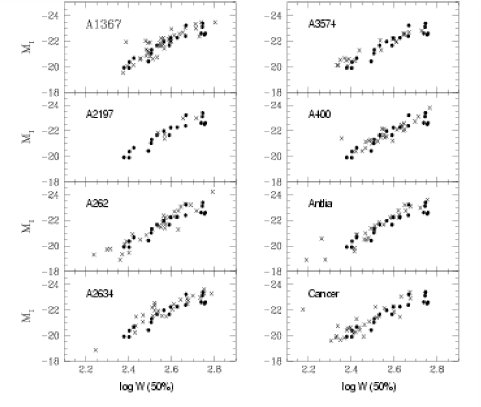}
\caption{Tully-Fisher distances to clusters of galaxies yield the Hubble Constant.}
\end{figure}

Next comes the fundamental plane. Kelson \etal~ (2000) used Cepheid distances to
the Leo, Virgo, and Fornax clusters to calibrate the fundamental plane and the
D$_n$,$\sigma$ relation (Figure 20), obtaining H$_0$ = 78 $\pm$ 5 $\pm$ 9 km s$^{-1}$ Mpc$^{-1}$.

\begin{figure}[ht]
\includegraphics[width=10cm]{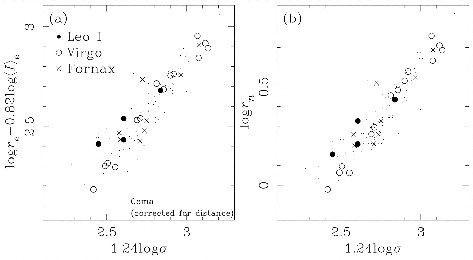}
\caption{Calibration of the fundamental plane (left) and the D$_n$,$\sigma$ relation (right).
This is superposed on the Coma cluster (dots).}
\end{figure}
\vfill\break
Type Ia supernovae were calibrated by Gibson \etal~ (2000), using 6 supernova hosts
with measured decline rates, some of them reworked from Saha \etal~ (1999). Figure 21
shows the application of the calibration to supernovae out to 30,000 km s$^{-1}$,
yielding H$_0$ = 71 $\pm$ 2 $\pm$ 7 km s$^{-1}$ Mpc$^{-1}$.

\begin{figure}[ht]
\includegraphics[width=5cm]{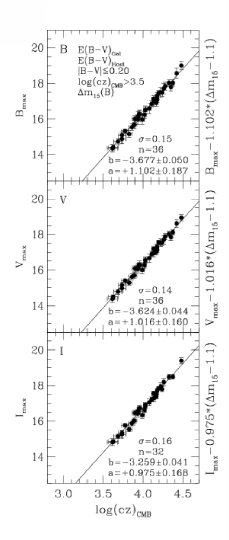}
\caption{Decline rate corrected supernovae in B,V,I bandpasses showing a dispersion of 0.16 mag.}
\end{figure}

Finally, Ferrarese \etal~ (2000a) calibrated surface brightness fluctuations in early type
galaxies, obtaining H$_0$ = 69 $\pm$ 4 $\pm$ 6 km s$^{-1}$ Mpc$^{-1}$.

A good summary of the results is provided by Figure 22, which shows the Cepheids and the calibrated 
distance indicators to a redshift of 0.1, well beyond the effect of local velocity field
perturbations. Hubble's (1929) distances are confined to the first tick mark in Figure 23.
\clearpage
\begin{figure}[ht]
\includegraphics[width=7.5cm]{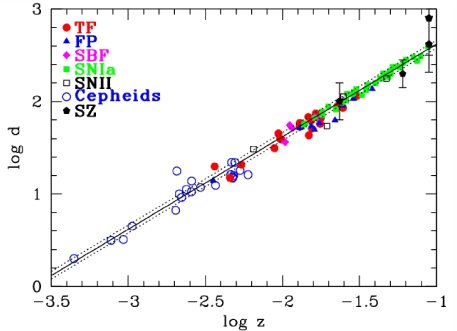}
\caption{Redshifts and distances for Cepheids, Tully-Fisher clusters, surface brightness fluctuations,
the fundamental plane, supernovae of type Ia and type II and clusters with Sunyaev-Zeldovich distances.}
\end{figure}
\begin{figure}[ht]
\includegraphics[width=15cm]{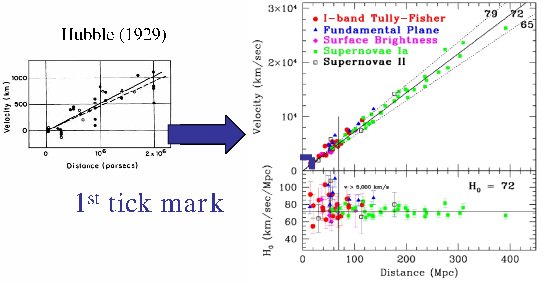}
\caption{The combined distance indicators yield H$_0$ = 72 km s$^{-1}$ Mpc$^{-1}$.}
\end{figure}

\vfill\break

The uncertainties in the Hubble Constant remain dominated by systematic errors. The first of these
is reddening, and this has been tested in work by Macri \etal~ (2001), who reobserved many of the
Key Project Cepheids with the NICMOS infrared camera. Their results for M81 are shown in Figure 24.
Overall, the H band distances agreed with the Key Project distances to 1\%.

\begin{figure}[ht]
\includegraphics[width=7.5cm]{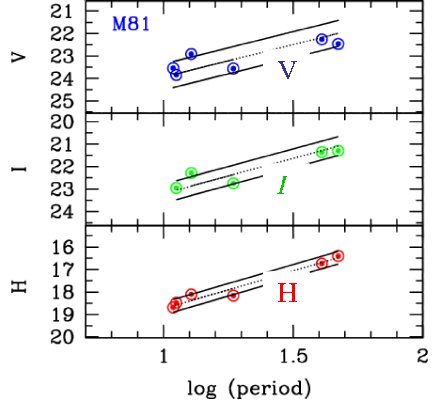}
\caption{A NICMOS H band period luminosity relation for M81 is compared with Key Project results.}
\end{figure}

We have dealt with the second systematic error, metallicity differences with the Large Magellanic
Cloud (LMC), but the LMC remains the principal systematic error because of the assumption of its
distance modulus, 18.50. A second fiducial distance has subsequently become available in the
maser distance of NGC 4258. Herrnstein \etal~ (1999) obtained the distance of this galaxy by fitting
a simple kinematic model to the maser radial velocities and VLBI proper motions (7.3 $\pm$ 0.4 Mpc).
The Cepheid distance is 7.5 $\pm$ 0.3 Mpc (Macri \etal~ 2006). In addition, HST trigonometric
parallaxes for a sample of Cepheids have become available (Benedict \etal~ 2007), confirming the
Key Project period-luminosity calibration, as shown in Figure 25.
\begin{figure}[ht]
\includegraphics[width=7.5cm]{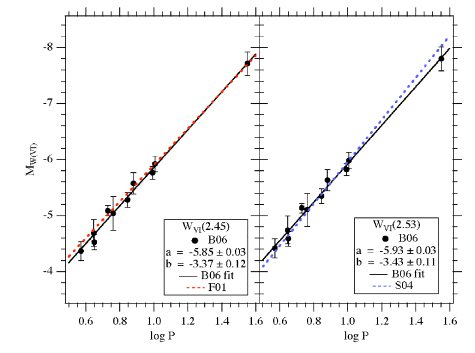}
\caption{HST parallaxes for Galactic Cepheids yield the period-luminosity relation shown by the
filled circles. The Freedman \etal~ (2001) PL relation is shown on the left and the Sandage \etal~
(2004) PL relation is shown on the right. The former is a better fit for the periods in excess
of 10 days which were used in the Key Project.}
\end{figure}

The probability distribution for H$_0$, combining the results of the secondary distance indicators
and the full error budget of the Key Project is shown in Figure 26 (Freedman \etal~ 2001).

\begin{figure}[ht]
\includegraphics[width=7.5cm]{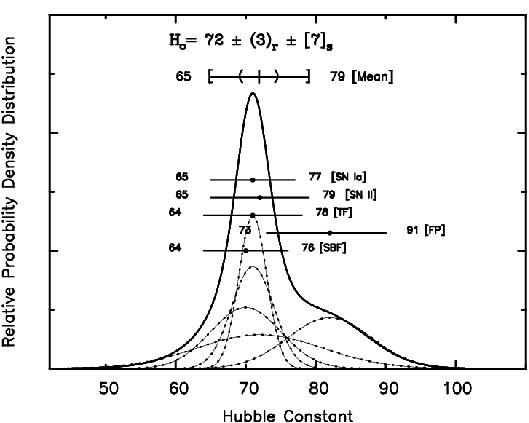}
\caption{The probability distribution emerging from the Key Project error analysis.}
\end{figure}

\section{The Team}
\noindent
Most of the team members (many referenced above) are captured in Figure 27. Special roles were
played by the individuals depicted in Figure 28 and 29.

\begin{figure}[ht]
\includegraphics[width=15cm]{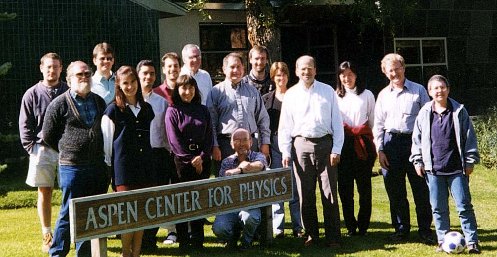}
\caption{The team photographed at the Aspen Center for Physics in 1999. From
left to right: Brad Gibson, Peter Stetson, Shaun Hughes, Laura Ferrarese,
Lucas Macri, Dan Kelson, Wendy Freedman, John Graham, Rob Kennicutt, John 
Huchra, Kim Sebo, Jeremy Mould, Shoko Sakai, Garth Illingworth and Nancy
Silbermann}
\end{figure}
\vfill\break
\begin{figure}[ht]
\includegraphics[width=10cm]{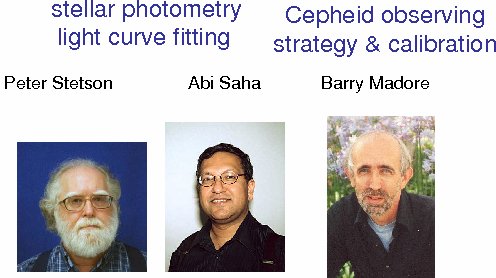}
\caption{Special roles were played by three team members.}
\end{figure}

\begin{figure}[ht]
\includegraphics[width=10cm]{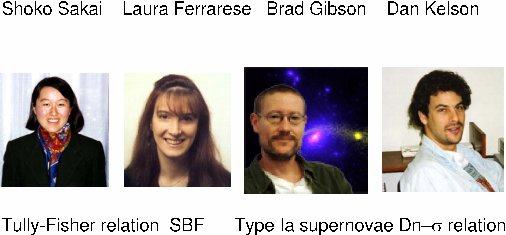}
\caption{Final papers on particular secondary distance indicators were written by four team members.}
\end{figure}

Other notable contributions were made by John Graham, Nancy Silbermann, Randy Phelps, Daya Rawson,
Fabio Bresolin, Lucas Macri, Bob Hill, Kim Sebo, Paul Harding, Anne Turner, Han Ming Sheng,
Shaun Hughes, Charles Prosser, John Huchra, Holland Ford, and Garth Illingworth. Jim Gunn, Sandra
Faber and John Hoessel were instrument team liaisons. The team drew on work from a large number
of individuals, including Brent Tully, Riccardo Giovanelli, Mario Hamuy, Mark Phillips, Bob Schommer,
Martha Haynes, John Tonry, Adam Riess, Bob Kirshner, Brian Schmidt, Gustav Tammann, Allan Sandage,
Mike Pierce, John Blakeslee, George Jacoby, Robin Ciardullo, Sandra Faber, Donald Lynden-Bell,
Gary Wegner, David Burstein, Alan Dressler, Roberto Terlevich, Roger Davies and Gerard de
Vaucouleurs.

\section{Subsequent Observations and Future Prospects}
\noindent
A number of important observations have been made since the publication of the Key Project results.
The Supernova and H$_0$ for the Equation of State (SHOES) project has carried out a differential
analysis of the supernova hosts NGC 4536, 4639, 3982, 3370, 3021 and 1309 with respect to NGC 4258.
This eliminates uncertainties such as photometric transformations and crucially, the LMC distance.
The maser distance to NGC 4258 is assumed instead. In this way, Riess \etal~ (2009) obtain
H$_0$ = 74.2 $\pm$ 3.6 km s$^{-1}$ Mpc$^{-1}$. They also find a value of w in the equation of state
w = P/$\rho$ = 1.12 $\pm$ 0.12.

Most importantly, the Hubble Constant has been deduced from the position of the first acoustic peak
in the small scale anisotropy of the cosmic microwave background. Given the sound horizon on the 
surface of last scattering (143 $\pm$ 4 Mpc) and the angular size of the first acoustic peak
(0.601 $\pm$ 0.005), one obtains an angular diameter distance for the surface of last scattering
of 13.7 $\pm$ 0.4 Gpc. Assuming $\Omega_M$ = 0.3 and $\Omega_\Lambda$ = 0.7, this yields a value
of H$_0$ of 70 km s$^{-1}$ Mpc$^{-1}$. Solving for all the cosmological model parameters,
Komatsu \etal~ (2009) find H$_0$ = 70.5 $\pm$ 1.3 km s$^{-1}$ Mpc$^{-1}$. 

Progress will continue in the classical distance scale of the Key Project too. The NASA mission
SIM-Lite is expected to yield a rotational parallax from velocities and proper motions for
the galaxies M31 and M33 to 1\% accuracy. This will provide a definitive Cepheid period-luminosity
relation.

\leftline{\bf Acknowledgements}
We wish to thank NASA for designating this a Key Project for the Hubble Space Telescope.
We wish to thank the team for making it a success. Finally, we wish to thank the Gruber
Foundation for recognizing this work.

\end{document}